\documentclass[fleqn,10pt]{wlscirep}
\usepackage[utf8]{inputenc}
\usepackage[T1]{fontenc}
\usepackage{tabularx}
\usepackage{seqsplit}
\title{Elucidating mechanism of optical cavities in superconducting strip single photon detectors using transmission line and impedance models}

\author[1,*]{Hiroki Kutsuma}
\author[1]{Taro Yamashita}
\affil[1]{Graduate School of Engineering, Tohoku University, Sendai, Miyagi 980-8579, Japan}

\affil[*]{hiroki.kutsuma.c5@tohoku.ac.jp}

\begin{abstract}
We clarified the physical mechanism of superconducting strip single photon detectors~(SSPDs) with optical cavities by using transmission line and impedance models.
By introducing the transmission line model, we derived the analytical formulae for the absorptance of SSPDs with optical cavities.
We compared the absorptance obtained from the analytical formulae for SSPDs with single-side, double-side, and dielectric multi-layer optical cavities against the results of numerical simulations.
The comparison showed that the results were nearly identical.
By introducing the impedance model, it was clearly shown that the SSPDs with optical cavities achieved the maximum absorptance when their input impedance of the SSPDs with optical cavities matched the impedance of the input medium.
The design concepts proposed in this study are applicable to other superconducting detectors, such as microwave kinetic inductance detectors and transition-edge sensors.
\end{abstract}
\begin{document}

\flushbottom
\maketitle
\thispagestyle{empty}

%\noindent Please note: Abbreviations should be introduced at the first mention in the main text – no abbreviations lists. Suggested structure of main text (not enforced) is provided below.

\section*{Introduction}

Superconducting strip single photon detectors~(SSPDs)~\cite{Goltsman:01} are widely used in quantum computing~\cite{zhong2020quantum}, quantum key distribution~\cite{sasaki2011field}, and quantum optics experiments~\cite{kobayashi2016frequency, endo2021quantum} because they offer high detection efficiencies~\cite{marsili2013detecting, zhang2017nbn, reddy2020superconducting, hu2020detecting, chang2021detecting, xu2021superconducting, reddy2022broadband}, low timing jitters~\cite{miki2018superconducting,korzh2020demonstration, colangelo2023impedance}, low dark count rates~\cite{yang2014superconducting,shibata2015ultimate,zhang2018fiber, chiles2022new}, and high count rates~\cite{zhang201916, resta2023gigahertz, craiciu2023high, hao2024compact}.
Among these properties, detection efficiency is the most critical parameter for photon detectors.
The system detection efficiency~(SDE) is defined as ${\rm SDE} = P_{\rm pulse}\times P_{\rm couple}\times P_{\rm abs}\times(1 - P_{\rm loss})$, where $P_{\rm pulse}$ is the pulse generation efficiency, $P_{\rm couple}$ is the coupling efficiency, $P_{\rm abs}$ is the absorptance, and $P_{\rm loss}$ denotes other losses. 
With various advancements~\cite{verevkin2002detection, rosfjord2006nanowire, kerman2007constriction, miki2009compactly, miki2010multichannel,gaggero2010nanowire,miller2011compact, clem2011geometry,marsili2013detecting,miki2013high}, SSPDs have achieved SDEs approaching unity at telecommunication wavelengths~\cite{reddy2020superconducting, hu2020detecting, chang2021detecting,reddy2022broadband}, offering a significantly higher detection efficiency than that of semiconductor detectors, such as avalanche photodiodes~(APDs)~\cite{fang2020ingaas} and frequency upconversion detectors~\cite{kamada2008efficient}. 

The introduction of optical cavities~\cite{rosfjord2006nanowire} has played a key role in significantly enhancing the absorptance $P_{\rm abs}$. 
Optical cavities store incident photons and enable efficient interaction with the superconducting wire.
The geometry of the optical cavities has been optimized through numerical simulations~\cite{anant2008optical,yamashita2013low,zhang2017nbn}, such as rigorous coupled-wave analysis~(RCWA)~\cite{moharam1981rigorous} or finite element method~(FEM)~\cite{turner1956stiffness}.
The simulation results reveal that the maximum absorptance strongly depends on the thickness and line-to-space ratio of the superconducting wire~\cite{yamashita2013low}.
Furthermore, it was found that in optical cavities with metallic mirrors, the thickness of the dielectric layer for achieving the maximum absorptance does not necessarily match the quarter-wave thickness of the incident photon wavelength scaled by its refractive index.
In~\citeonline{yamashita2013low}, the absorptance of the SSPDs with optical cavities was evaluated for various filling factors and their absorptance was simulated using a commercial FEM solver.
The experimentally measured absorptance tended to follow that predicted by the simulation.
This indicates that the simulations provide insight into the absorptance of SSPDs with optical cavities.
However, the underlying physical mechanism of SSPDs with optical cavities, by which they achieve the desired performance in simulations, remains unclear.
To address this gap, several analytical approaches have been proposed.
In \citeonline{driessen2009impedance}, the impedance model was introduced to analytically calculate the absorptance as a function of the thickness of a superconducting wire sandwiched between two dielectric layers with infinite thicknesses.
The maximum absorptance depends only on the refractive indices of the two dielectric layers, whereas the required thickness of the superconducting wire to achieve the maximum absorptance depends only on the refractive indices of the two dielectric layers, the superconducting material, slit material, and filling factor.
More recently, the transmission line model, which is commonly used in the design of microwave circuits, was applied to calculate the absorptance of the optical cavities in microwave kinetic inductance detectors~(MKIDs)~\cite{kouwenhoven2022model}.
However, these approaches are not applicable to the calculation of SSPDs with optical cavities, and do not provide the physical mechanism for achieving the maximum absorptance of such SSPDs.

In this study, we applied the transmission line and impedance models~\cite{pozar2011microwave} to the design of SSPDs with optical cavities to provide design guidelines with physical considerations.
We propose a design methodology for SSPDs with optical cavities and derive the analytical formulae for the thicknesses of the superconducting wire and dielectric layer required to achieve the maximum absorptance. 
Finally, we clarify the physical mechanism underlying the maximum absorptance.

\section*{Transmission line model for SSPDs}

To apply the transmission line model to the design of SSPDs with optical cavities, we impose the following conditions: The wavelength of the incident photon is much longer than the wire width, such as in superconducting nanostrip single photon detectors~(SNSPDs), or much shorter than the wire width, such as in superconducting wide strip single photon detectors~(SWSPDs)~\cite{yabuno2023superconducting}~(Condition~I). 
For the calculation of SNSPDs, the geometry is assumed to be a meander structure~\cite{verevkin2002detection}, and the polarization of the incident photon is aligned parallel to the wire axis~(Condition~II).
The loss in the dielectric layer is negligible~(Condition~III).
The relative permeabilities of both the metal and dielectric layers are equal to 1~(Condition~IV).
The imaginary part of the refractive index of the mirror is much larger than its real part, as in the case with commonly used materials such as silver and gold in the optical cavities~\cite{marsili2013detecting, miki2013high,yamashita2013low,chang2021detecting}~(Condition~V).
The thickness of the mirror is sufficient to prevent the penetration of the incident photons~(Condition~VI).
The superconducting wire is much thinner than the wavelength of the incident photon~(Condition~VII).
In this paper, the impedance, wavelength, and wave number of vacuum are denoted as $\eta_0$, $\lambda_0$, and $k_0$, respectively. 
We also define the complex relative permittivity $\varepsilon _{\rm M}$~(complex refractive index $n_{\rm M}$) of a metal as $\varepsilon_{\rm M} = \varepsilon_{\rm Mr} - i\varepsilon_{\rm Mi}$~($n_{\rm M} = n_{\rm Mr} - in_{\rm Mi}$), where $\varepsilon_{\rm Mr}$~($n_{\rm Mr}$) and $\varepsilon_{\rm Mi}$~($n_{\rm Mi}$) represent the real and imaginary parts of the complex relative permittivity~(complex refractive index) of the metal.
Note that $\varepsilon_{\rm Mr}$, $\varepsilon_{\rm Mi}$, $n_{\rm Mr}$, and $n_{\rm Mi}$ are positive values and the complex relative permittivity is related to the refractive index given by $n_{\rm M} = \sqrt{\varepsilon_{\rm M}}$.

The structure of SSPDs with optical cavities is regarded as a stacked arrangement consisting of the superconducting wire layer, dielectric layer, and metallic mirror layer, sandwiched between the input and output media.
The $F$ matrix of the layer ${\rm x}$ is given by the following formula~\cite{pozar2011microwave, kouwenhoven2022model}:
\begin{equation}
F_{\rm x} = 
\begin{pmatrix}
    F_{{\rm x}, 11}&F_{{\rm x}, 12}\\
    F_{{\rm x}, 21}&F_{{\rm x}, 22}
\end{pmatrix}
= 
\begin{pmatrix}
    \cosh(\gamma_{\rm x}d_{\rm x})&\eta_{\rm x}\sinh(\gamma_{\rm x}d_{\rm x})\\
    \frac{1}{\eta_{\rm x}}\sinh(\gamma_{\rm x}d_{\rm x})&\cosh(\gamma_{\rm x}d_{\rm x})
\end{pmatrix},
\end{equation}
where $d_{\rm x}$, $\eta_{\rm x}$, and $\gamma_{\rm x}$ are the thickness, characteristic impedance, and propagation constant of the layer~x, respectively. 
Here, the subscript~x may represent the wire layer ~(${\rm w}$), dielectric layer~(${\rm c}$), or metallic mirror layer~(${\rm m}$).
In Condition~IV, $\eta_{\rm x}$ and $\gamma_{\rm x}$ are defined as $\eta_{\rm x} = \eta_0/n_{\rm x}$ and $\gamma_{\rm x} = ik_0n_{\rm x}$, where $n_{\rm x}$ is the complex refractive index of the layer~x.
Similar to the transmission line of microwave circuits, the $F$ matrix of the total structure $F_{\rm t}$ is given by the product of each matrix from the input to output media.
The absorptance of SSPDs with optical cavities is given by
\begin{equation}
    A = 1 - \bar{r}r - \bar{t}t,
\label{eq:absorptance}
\end{equation}
where $r$ and $t$ are the reflection and transmission coefficients, respectively, and $\bar{r}$ and $\bar{t}$ are their complex conjugates.
In \citeonline{frickey1994conversions}, $r$ and $t$ are expressed as
\begin{equation}
    r = \frac{F_{\rm t, 11}\eta_{\rm o} + F_{\rm t, 12} - F_{\rm t, 21}\bar{\eta_{\rm i}}\eta_{\rm o} - F_{\rm t, 22}\bar{\eta_{\rm i}}}{F_{\rm t, 11}\eta_{\rm o} + F_{\rm t, 12} + F_{\rm t, 21}\eta_{\rm i}\eta_{\rm o} + F_{\rm t, 22}\eta_{\rm i}}
\label{eq:refractance}
\end{equation}
and
\begin{equation}
    t = \frac{2\sqrt{{\rm Re}(\eta_{\rm i}){\rm Re}(\eta_{\rm o})}}{F_{\rm t, 11}\eta_{\rm o} + F_{\rm t, 12} + F_{\rm t, 21}\eta_{\rm i}\eta_{\rm o} + F_{\rm t,22}\eta_{\rm i}},
\label{eq:transmitance}
\end{equation}
where $\eta_{\rm i}$ and $\eta_{\rm o}$ are the impedances of the input and output media and $\bar{\eta_{\rm i}}$ represents the complex conjugate of ${\eta_{\rm i}}$.
From Condition~VI, the second term of the right-hand side ($\bar{t}t$) in Eq.~(\ref{eq:absorptance}) is negligible.
By satisfying Condition~I and Condition~II, the complex relative permittivity of the wire layer $\varepsilon_{\rm w}$, which is defined as the layer included in the superconducting wire and slit dielectric, is given by~\cite{aspnes1982local, driessen2009impedance}
\begin{equation}
    \varepsilon_{\rm w} = \varepsilon_{\rm metal}f + \varepsilon_{\rm slit}(1 - f),
\label{eq:epsilonw}
\end{equation}
where $\varepsilon_{\rm metal}$ and $\varepsilon_{\rm slit}$ are the complex relative permittivities of the metal and slit dielectric, respectively, and $f$ is the filling factor, which is the ratio of the line width to the sum of line and slit widths. 
Because the superconducting wire is much thinner than the wavelength of the incident photon~(Condition~VII), the $F$ matrix of the wire layer, $F_{\rm w}$, is approximated by
\begin{equation}
    F_{\rm w} \approx
    \begin{pmatrix}
        1&\eta_{\rm w}\gamma_{\rm w}d_{\rm w}\\
        \gamma_{\rm w}d_{\rm w}/\eta_{\rm w}&1
    \end{pmatrix},
    \label{eq:Fw}
\end{equation}
where $\gamma_{\rm w}$, $\eta_{\rm w}$, and $d_{\rm w}$ are the propagation constant, characteristic impedance, and thickness of the wire layer, respectvely.

\begin{figure}[t]
\centering\includegraphics[width=13cm]{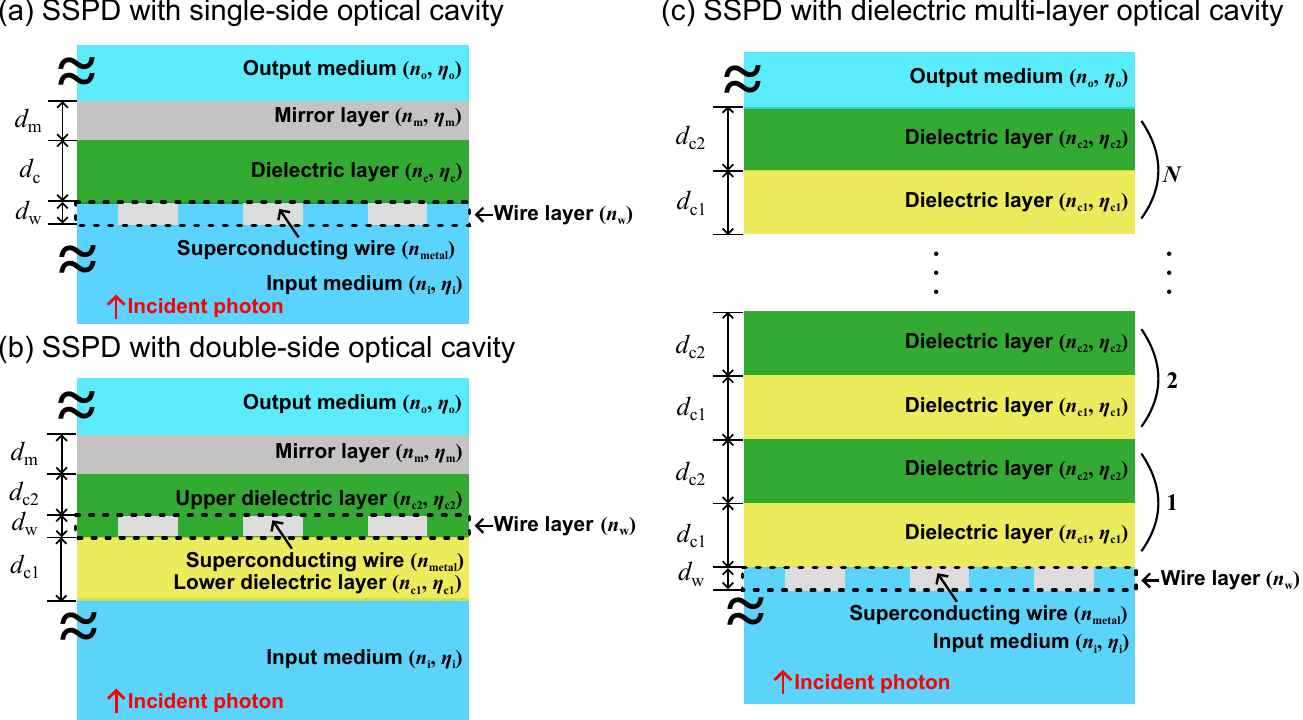}
\caption{Schematics of SSPDs with optical cavities. (a)~SSPD with single-side optical cavity. (b)~SSPD with double-side optical cavity. (c)~SSPD with dielectric multi-layer optical cavity. The two dielectric layers with different refractive indices are stacked in $N$ periods.}
\label{fig:photon_cavity_type}
\end{figure}

Because we have set the dielectric loss in the dielectric layer as negligible (Condition~III) and its thickness is commonly considered as nearly one-quarter of the incident photon wavelength scaled by its refractive index, the $F$ matrix of the dielectric layer $F_{\rm c}$ is approximated by
\begin{equation}
    F_{\rm c} \approx
    \begin{pmatrix}
        -\Delta\phi_{\rm c}&i\eta_{\rm c}\\
        i/\eta_{\rm c}&-\Delta\phi_{\rm c}
    \end{pmatrix}.
    \label{eq:Fc}
\end{equation}
$\Delta\phi_{\rm c}$ is defined as $\phi_{\rm c} = \pi/2 + \Delta\phi_{\rm c}$, where $\phi_{\rm c} = k_0n_{\rm c}d_{\rm c}$, and $n_{\rm c}$ and $d_{\rm c}$ are the refractive index and thickness of the dielectric layer, respectively.
Note that $\Delta\phi_{\rm c}$ is equal to $0$ when the thickness of the dielectric layer is one-quarter of the incident photon wavelength scaled by its refractive index.
In Condition~VI, the output medium is replaced by a metallic mirror, i.e., the output medium is treated as a metallic mirror with infinite thickness.
Therefore, the characteristic impedance of the output medium, $\eta_{\rm o}$, is replaced with $\eta_{\rm o}\approx i{\rm Im}(\eta_{\rm m})$. 
This approximation is based on the fact that the imaginary part of the refractive index of the metallic mirror is much larger than its real part~(Condition~V).
Using the above equations and approximations, we apply the transmission line model to SSPDs with single-side optical cavities~(Fig.~\ref{fig:photon_cavity_type}~(a)), double-side optical cavities~(Fig.~\ref{fig:photon_cavity_type}~(b)), and dielectric multi-layer optical cavities~(Fig.~\ref{fig:photon_cavity_type}~(c)).
All these configurations are reported to achieve high system detection efficiencies~\cite{marsili2013detecting, zhang2017nbn, reddy2020superconducting, hu2020detecting, chang2021detecting, xu2021superconducting,reddy2022broadband}.

\subsection*{Single-side optical cavity}

Here, we apply the transmission line model to SSPDs with single-side optical cavities, which consist of a dielectric layer with a metallic mirror placed on the superconducting wire, as shown in Fig.~\ref{fig:photon_cavity_type}~(a). First, to derive the thickness of the wire layer that achieve the maximum absorptance, we simplify the material of the mirror as a perfect electrical conductor~(PEC), i.e., $\eta_{\rm m} = 0$, and the thickness of the dielectric layer as one-quarter of the incident photon wavelength scaled by its refractive index, i.e., $\Delta\phi_{\rm c} = 0$.
Based on the approximations given in Eq.~(\ref{eq:Fw}) and Eq.~(\ref{eq:Fc}), the total $F$ matrix of the SSPDs with single-side optical cavities, $F_{\rm SSC}$, is approximated by
\begin{equation}
    F_{\rm SSC} = F_{\rm w}\times F_{\rm c} =
    \begin{pmatrix}
        i\eta_{\rm w}\gamma_{\rm w}d_{\rm w}/\eta_{\rm c}&i\eta_{\rm c}\\ i/\eta_{\rm c}&i\eta_{\rm c}\gamma_{\rm w}d_{\rm w}/\eta_{\rm w}
    \end{pmatrix}.
\label{eq:FSSC}
\end{equation}
By substituting Eq.~(\ref{eq:FSSC}) into Eq.~(\ref{eq:absorptance}) and~Eq.~(\ref{eq:refractance}), the absorptance of the SSPDs with single-side optical cavities, $A_{\rm SSC}$, is given by
\begin{equation}
    A_{\rm SSC} = \frac{-4k_0{\rm Im}(\varepsilon_{\rm w})n_{\rm i}d_{\rm w}}{(n_{\rm i} - k_0{\rm Im}(\varepsilon_w)d_{\rm w})^2 +~(k_0{\rm Re}(\varepsilon_{\rm w})d_{\rm w})^2},
\label{eq:ASSC}
\end{equation}
where $n_{\rm i}$ is the refractive index of the input medium.
An important point to note is that the absorptance is characterized only by the refractive index of the input medium, complex relative permittivity, and thickness of the wire layer.
The maximum absorptance $A_{\rm SSC}^{\rm max}$ and thickness of the wire layer at the maximum absorptance, $d^{\rm max}_{\rm w, SSC}$,  satisfy ${\rm d} A_{\rm SSC}/{\rm d}d_{\rm w} = 0$, resulting in
\begin{equation}
    d^{\rm max}_{\rm w, SSC} = \frac{n_{\rm i}}{k_0|\varepsilon_{\rm w}|}
    \label{eq:dSSCmax}
\end{equation}
and
\begin{equation}
    A_{\rm SSC}^{\rm max} = \frac{2{\rm Im}(\varepsilon_{\rm w})}{{\rm Im}(\varepsilon_{\rm w}) - |\varepsilon_{\rm w}|}.
\label{eq:AmaxSSC}
\end{equation}
As a result, we found that the required thickness of the wire layer to achieve the maximum absorptance was characterized by the refractive index of the input medium and complex relative permittivity of the wire layer, whereas $A_{\rm SSC}^{\rm max}$ is characterized by the complex relative permittivity of the wire layer.
The results indicate that the thickness of the wire layer at the maximum absorptance depends on the filling factor, because the relative permittivity of the wire layer depends on the filling factor, as shown in Eq.~(\ref{eq:epsilonw}).

Next, we derive the thickness of the dielectric layer at the maximum absorptance.
In this calculation, we fix the thickness of the wire layer at the maximum absorptance determined by Eq.~(\ref{eq:dSSCmax}). 
Here, we assume the metallic mirror, i.e., $\eta_{\rm m} \neq 0$.
In this condition, the $F$ matrix is approximated by
\begin{equation}
    F'_{\rm SSC} = 
    \begin{pmatrix}
        -\Delta\phi_{\rm c} + i\eta_{\rm w}\gamma_{\rm w}d^{\rm max}_{\rm w, SSC}/\eta_{\rm c}&i\eta_{\rm c}\\
        i/\eta_{\rm c}&-\Delta\phi_{\rm c} + i\eta_{\rm c}\gamma_{\rm w}d_{\rm w, SSC}^{\rm max}/\eta_{\rm w}
    \end{pmatrix}.
    \label{eq:FSSCdash}
\end{equation}
Here, we ignore the product of $d^{\rm max}_{\rm w, SSC}$ and $\Delta\phi_{\rm c}$ in this calculation, as it is sufficiently small.
By substituting Eq.~(\ref{eq:FSSCdash}) into Eq.~(\ref{eq:absorptance}) and~Eq.~(\ref{eq:refractance}), the absorptance $A'_{\rm SSC}$ is given by
\begin{equation}
    A'_{\rm SSC} = \frac{-4{\rm Im}(\varepsilon_{\rm w})/|\varepsilon_{\rm w}|}{(1 - {\rm Im}(\varepsilon_{\rm w})/|\varepsilon_{\rm w}|)^2 + ({\rm Re}(\varepsilon_{\rm w})/|\varepsilon_{\rm w}| - n_{\rm c}^2{\rm Im}(n_{\rm m})/|n_{\rm m}|^2n_{\rm i} + \Delta\phi_{\rm c}n_{\rm c}/n_{\rm i})^2}.
\label{eq:ASSC2}
\end{equation}
Here, we ignore the product of ${\rm Im}(\eta_{\rm m})\cdot d^{\rm max}_{\rm w, SSC}$ and ${\rm Im}(\eta_{\rm m})\cdot\Delta\phi_{\rm c}$ because the imaginary part of the refractive index of the metallic mirror is large, implying that the imaginary part of the impedance of the metallic mirror is small.
The thickness of the dielectric layer at the maximum absorptance, $d_{\rm c, SSC}^{\rm max}$, and the maximum absorptance $A_{\rm SSC}^{'\rm max}$  are given by
\begin{equation}
    d_{\rm c, SSC}^{\rm max} = \frac{\lambda_0}{4n_{\rm c}} - \frac{{n_{\rm i}\rm Re}(\varepsilon_{\rm w})}{k_0n_{\rm c}^2|\varepsilon_{\rm w}|} + \frac{{\rm Im}(n_{\rm m})}{k_0|n_{\rm m}|^2}
\label{eq:dSSCcmax}
\end{equation}
and
\begin{equation}
    A_{\rm SSC}^{'\rm max} = \frac{-4{\rm Im}(\varepsilon_{\rm w})}{|\varepsilon_{\rm w}|(1 - {\rm Im}(\varepsilon_{\rm w})/|\varepsilon_{\rm w}|)^2}.
\label{eq:AmaxSSC2}
\end{equation}
As a result, it was found that the thickness of the dielectric layer at the maximum absorptance was less than one-quarter of the incident photon wavelength scaled by its refractive index.
This is because of the real part of the relative permittivity of the wire layer and imaginary part of the refractive index of the metal.

\subsection*{Double-side optical cavity}

As shown in Fig.~\ref{fig:photon_cavity_type}~(b), we apply the transmission line model to SSPDs with double-side optical cavities. 
These SSPDs have dielectric layers placed both above and below the superconducting wire, and a metallic mirror is placed on top of the upper dielectric layer.
First, to derive the thickness of the wire layer to achieve the maximum absorptance, we used a PEC mirror as the metallic mirror and set the thicknesses of the upper and lower dielectric layers as one-quarter of the incident photon wavelength scaled by their refractive indices.
The $F$ matrix of the SSPDs with double-side optical cavities is simplified as
\begin{equation}
    F_{\rm DSC}  = F_{\rm c1}\times F_{\rm w}\times F_{\rm c2} =
    \begin{pmatrix}
        -\eta_{\rm c1}/\eta_{\rm c2}& -\eta_{\rm c1}\eta_{\rm c2}\gamma_{\rm w}d_{\rm w}/\eta_{\rm w}\\
        -\eta_{\rm w}\gamma_{\rm w}d_{\rm w}/\eta_{\rm c1}\eta_{c2}&-\eta_{\rm c2}/\eta_{\rm c1}
    \end{pmatrix},
\label{eq:FDSC}
\end{equation}
where the subscripts ${\rm c1}$ and ${\rm c2}$ denote the lower and upper dielectric layers, respectively, as shown in Fig.~\ref{fig:photon_cavity_type}~(b).
We used a similar approach for SSPDs with single-side optical cavities to derive the dependence of the absorptance ($A_{\rm DSC}$) on the thickness of the wire layer, thickness of the wire layer at the maximum absorptance $d_{\rm w, DSC}^{\rm max}$, and the maximum absorptance $A_{\rm DSC}^{\rm max}$.
$A_{\rm DSC}$,  $d_{\rm w, DSC}^{\rm max}$, and $A_{\rm DSC}^{\rm max}$ are given by
\begin{equation}
    A_{\rm DSC} = \frac{-4k_0n_{\rm i}d_{\rm w}{\rm Im}(\varepsilon_{\rm w})/n_{c1}^2}{(k_0n_{\rm i}d_{\rm w}{\rm Im}(\varepsilon_{\rm w})/n_{\rm c1}^2 - 1)^2 +~(k_0n_{\rm i}d_{\rm w}{\rm Re}(\varepsilon_{\rm w})/n_{c1}^2)^2},
\label{eq:ADSC}
\end{equation}
\begin{equation}
    d_{\rm w, DSC}^{\rm max} = \frac{n_{c1}^2}{k_0n_{\rm i}|\varepsilon_{\rm w}|},
\label{eq:dDSCmax}
\end{equation}
and
\begin{equation}
    A_{\rm DSC}^{\rm max} = \frac{2{\rm Im}(\varepsilon_{\rm w})}{{\rm Im}(\varepsilon_{\rm w}) - |\varepsilon_{\rm w}|}.
\label{eq:AmaxDSC}
\end{equation}
The results indicate that $d_{\rm w, DSC}^{\rm max}$ depends on not only the refractive index of the input medium and complex relative permittivity of the wire layer but also the refractive index of the lower dielectric layer.
The maximum absorptance of the SSPDs with double-side optical cavities, $A_{\rm DSC}^{\rm max}$, is the same as that of SSPDs with single-side optical cavities, as shown in Eq.~(\ref{eq:AmaxSSC}).

Next, to derive the thickness of the lower and upper dielectric layers to achieve the maximum absorptance, we used the thickness of the wire layer given by Eq.~(\ref{eq:dDSCmax}) and applied the characteristic of the metallic mirror, i.e., $\eta_{\rm m}\neq0$.
The $F$ matrix of SSPDs with double-side optical cavities is given by
\begin{equation}
    F'_{\rm DSC} =
    \begin{pmatrix}
        -\frac{\eta_{\rm c1}}{\eta_{\rm c2}}&-i(\eta_{\rm c2}\Delta\phi_{\rm c1} + \eta_{\rm c1}\Delta\phi_{\rm c2}) - \frac{\eta_{\rm c1}\eta_{\rm c2}\gamma_{\rm w}d_{\rm w}}{\eta_{\rm w}}\\
        -i\left(\frac{\Delta\phi_{\rm c2}}{\eta_{\rm c1}} + \frac{\Delta\phi_{\rm c1}}{\eta_{\rm c2}}\right) - \frac{\eta_{\rm w}\gamma_{\rm w}d_{\rm w}}{\eta_{\rm c1}\eta_{\rm c2}}&-\frac{\eta_{\rm c2}}{\eta_{\rm c1}}
    \end{pmatrix},
\end{equation}
where $\Delta\phi_{\rm c1}$ ($\Delta\phi_{\rm c2}$) is defined by $\phi_{\rm c1} = \pi/2 + \Delta\phi_{\rm c1}$ ($\phi_{\rm c2} = \pi/2 + \Delta\phi_{\rm c2}$), where $\phi_{\rm c1}=k_0n_{\rm c1}d_{\rm c1}$ ($\phi_{\rm c2}=k_0n_{\rm c2}d_{\rm c2}$), and $n_{\rm c1}$ ($n_{\rm c2}$) and $d_{\rm c1}$ ($d_{\rm c2}$) are the refractive index and thickness of the lower (upper) dielectric layer, respectvely.
Here, we ignore the products of $\Delta\phi_{c1}$, $\Delta\phi_{c2}$, and $d_{\rm w}$, as their values are small.
As in the derivation of the equations for SSPDs with single-side optical cavities, the absorptance $A'_{\rm DSC}$ is given by
\begin{equation}
    A'_{\rm DSC} = \frac{-4{\rm Im}(\varepsilon_{\rm w})/|\varepsilon_{\rm w}|}{(1 - {\rm Im}(\varepsilon_{\rm w})/|\varepsilon_{\rm w}|)^2 +~({\rm Re}(\varepsilon_{\rm w})/|\varepsilon_{\rm w}| + \Delta\phi_{\rm DSC}n_{\rm i}/n_{\rm c1} - n^2_{\rm c2}n_{\rm i}{\rm Im}(n_{\rm m})/n_{\rm c1}^2|n_{\rm m}|^2)^2},
\label{eq:ADSC2}
\end{equation}
where $\Delta\phi_{\rm DSC}$ is defined by
\begin{equation}
    \Delta\phi_{\rm DSC} = \Delta\phi_{\rm c1} + \frac{n_{\rm c2}}{n_{\rm c1}}\Delta\phi_{\rm c2}.
\label{eq:dphiDSC_dash}
\end{equation}
Therefore, $\Delta\phi_{\rm DSC}^{\rm max}$, which is the value of $\Delta\phi_{\rm DSC}$ at the maximum absorptance, and the maximum absorptance $A_{\rm DSC}^{'\rm max}$ are given by
\begin{equation}
    \Delta\phi_{\rm DSC}^{\rm max} = -\frac{n_{\rm c1}}{n_{\rm i}}\frac{{\rm Re}(\varepsilon_{\rm w})}{|\varepsilon_{\rm w}|} + \frac{n_{\rm c2}^2{\rm Im}(n_{\rm m})}{n_{\rm c1}|n_{\rm m}|^2}
\label{eq:dphiDSC}
\end{equation}
and
\begin{equation}
    A_{\rm DSC}^{'\rm max} = \frac{-4{\rm Im}(\varepsilon_{\rm w})}{|\varepsilon_{\rm w}|(1 - {\rm Im}(\varepsilon_{\rm w})/|\varepsilon_{\rm w}|)^2}.
\end{equation}
The results indicate that the maximum absorptance of SSPDs with double-side optical cavities is the same as that of single-side optical cavities, as shown in Eq.~(\ref{eq:AmaxSSC2}).
As shown in Eq.~(\ref{eq:dphiDSC_dash}) and Eq.~(\ref{eq:dphiDSC}), various combinations of the thicknesses of the lower and upper dielectric layers can yield the maximum absorptance, as $\Delta\phi_{\rm c1}$ and $\Delta\phi_{\rm c2}$ depend on the thickness of the upper and lower dielectric layers, respectively.

We present an example of the thickness combinations of the two dielectric layers that achieve the maximum absorptance.
When the thickness of the lower dielectric layer is one-quarter of the incident photon wavelength by its refractive index, i.e., $d_{\rm c1} = \lambda_0/4n_{\rm c1}$, the thickness of the upper dielectric layer that yields the maximum absorptance, $d_{\rm c2, DSC}^{\rm max}$, is given by
\begin{equation}
    d_{\rm c2, DSC}^{\rm max} = \frac{\lambda_0}{4n_{\rm c2}} - \frac{n_{\rm c1}^2{\rm Re}(\varepsilon_{\rm w})}{k_0n_{\rm i}n_{\rm c2}^2|\varepsilon_{\rm w}|} + \frac{{\rm Im}(n_{\rm m})}{k_0|n_{\rm m}|^2}.
\label{eq:dc2max}
\end{equation}
The result indicates that the thickness of the dielectric layer at the maximum absorptance is less than one-quarter of the incident photon wavelength scaled by its refractive index.
This is because of the real part of the relative permittivity of the wire layer and imaginary part of the refractive index of the metal.

\subsection*{Dielectric multi-layer optical cavity}

Here, we apply the transmission line model to SSPDs with dielectric multi-layer optical cavities, shown in Fig.~\ref{fig:photon_cavity_type}~(c).
The dielectric multi-layer optical cavities comprise two dielectric layers with different refractive indices, which are alternatively stacked on the superconducting wire.
The thickness of each layer is one-quarter of the incident photon wavelength scaled by its refractive index.

The $F$ matrix of two dielectric layers, indicated by subscripts ${\rm c1}$ and ${\rm c2}$, are given by
\begin{equation}
    F_{\rm c1} =
    \begin{pmatrix}
        0&i\eta_{\rm c1}\\
        i/\eta_{\rm c1}&0
    \end{pmatrix}
\end{equation}
and
\begin{equation}
    F_{\rm c2} =
    \begin{pmatrix}
        0&i\eta_{\rm c2}\\
        i/\eta_{\rm c2}&0
    \end{pmatrix}.
\end{equation}
The $F$ matrix of the two stacked dielectric layers with different refractive indices (one period), $F_{\rm c12}$, is given by
\begin{equation}
    F_{\rm c12} = F_{\rm c1}F_{\rm c2} =
    \begin{pmatrix}
        -\eta_{\rm c1}/\eta_{\rm c2}&0\\
        0&-\eta_{\rm c2}/\eta_{\rm c1}
    \end{pmatrix}.
\end{equation}
Therefore, the $F$ matrix for $N$ periods is given by
\begin{equation}
    F_{\rm c12}^N =
    \begin{pmatrix}
       ~(-\eta_{\rm c1}/\eta_{\rm c2})^N&0\\
        0&(-\eta_{\rm c2}/\eta_{\rm c1})^N
    \end{pmatrix}.
\end{equation}
The $F$ matrix of SSPDs with dielectric multi-layer optical cavities, $F_{\rm MLC}$, is given by
\begin{equation}
    F_{\rm MLC} = F_{\rm w}F_{\rm c12}^N =
    \begin{pmatrix}
       ~(-\eta_{\rm c1}/\eta_{\rm c2})^N&(-\eta_{\rm c2}/\eta_{\rm c1})^N\eta_{\rm w}\gamma_{\rm w}d_{\rm w}\\
       ~(-\eta_{\rm c1}/\eta_{\rm c2})^N\gamma_{\rm w}d_{\rm w}/\eta_{\rm w}&(-\eta_{\rm c2}/\eta_{\rm c1})^N
    \end{pmatrix}.
\label{eq:FML}
\end{equation}
Substituting Eq.~(\ref{eq:FML}) into Eq.~(\ref{eq:refractance}) and Eq.~(\ref{eq:transmitance}), the reflection coefficient $r_{\rm MLC}$ and transmission coefficient $t_{\rm MLC}$ of SSPDs with dielectric multi-layer optical cavities are given by
\begin{equation}
    r_{\rm MLC} = \frac{(-\eta_{\rm c1}/\eta_{\rm c2})^N\eta_{\rm o} +~(-\eta_{\rm c2}/\eta_{\rm c1})^N\eta_{\rm w}\gamma_{\rm w}d_{\rm w} -~(-\eta_{\rm c1}/\eta_{\rm c2})^N\gamma_{\rm w}\eta_{\rm o}\eta_{\rm i}/\eta_{\rm w} -~(-\eta_{\rm c2}/\eta_{\rm c1})^N\eta_{\rm i}}{(-\eta_{\rm c1}/\eta_{\rm c2})^N\eta_{\rm o} +~(-\eta_{\rm c2}/\eta_{\rm c1})^N\eta_{\rm w}\gamma_{\rm w}d_{\rm w} +~(-\eta_{\rm c1}/\eta_{\rm c2})^N\gamma_{\rm w}\eta_{\rm o}\eta_{\rm i}/\eta_{\rm w} +~(-\eta_{\rm c2}/\eta_{\rm c1})^N\eta_{\rm i}}
\end{equation}
and
\begin{equation}
    t_{\rm MLC} = \frac{2\eta_{\rm o}\eta_{\rm i}}{(-\eta_{\rm c1}/\eta_{\rm c2})^N\eta_{\rm o} +~(-\eta_{\rm c2}/\eta_{\rm c1})^N\eta_{\rm w}\gamma_{\rm w}d_{\rm w} +~(-\eta_{\rm c1}/\eta_{\rm c2})^N\gamma_{\rm w}\eta_{\rm o}\eta_{\rm i}/\eta_{\rm w} +~(-\eta_{\rm c2}/\eta_{\rm c1})^N\eta_{\rm i}}.
\end{equation}
Here, we ignore the imaginary part of the refractive indices of the input and output media.
By assuming that the refractive indices satisfy $n_{\rm c1} < n_{\rm c2}$, resulting in $\eta_{\rm c1} > \eta_{\rm c2}$, and $N$ is sufficiently large, i.e., $|-\eta_{\rm c1}/\eta_{\rm c2}|^N \gg |-\eta_{\rm c2}/\eta_{\rm c1}|^N$, the transmission coefficient becomes negligible.
The reflection coefficient can be approximated by
\begin{equation}
    r_{\rm MLC} = \frac{1 - \gamma_{\rm w}d_{\rm w}\eta_{\rm i}/\eta_{\rm w}}{1 + \gamma_{\rm w}d_{\rm w}\eta_{\rm i}/\eta_{\rm w}}.
    \label{eq:rMLC}
\end{equation}
By substituting Eq.~(\ref{eq:rMLC}) into Eq.~(\ref{eq:absorptance}), the absorptance of SSPDs with dielectric multi-layer optical cavities is given by
\begin{equation}
    A_{\rm MLC} = \frac{-4k_0{\rm Im}(\varepsilon_{\rm w})n_{\rm i}d_{\rm w}}{(n_{\rm i} - k_0{\rm Im}(\varepsilon_w)d_{\rm w})^2 +~(k_0{\rm Re}(\varepsilon_{\rm w})d_{\rm w})^2}.
\label{eq:AML}
\end{equation}
The result is the same as the absorptance of SSPDs with single-side optical cavities, as shown in Eq.~(\ref{eq:ASSC}).
It indicates that the thickness of the wire layer at the maximum absorptance, $d_{\rm MLC}^{\rm max}$, and the maximum absorptance $A_{\rm MLC}^{\rm max}$ are the same as those of SSPDs with single-side optical cavities given by
\begin{equation}
    d^{\rm max}_{\rm MLC} = \frac{n_{\rm i}}{k_0|\varepsilon_{\rm w}|}
\label{eq:dMLmax}
\end{equation}
and
\begin{equation}
    A_{\rm MLC}^{\rm max} = \frac{2{\rm Im}(\varepsilon_{\rm w})}{{\rm Im}(\varepsilon_{\rm w}) - |\varepsilon_{\rm w}|}.
\label{eq:AmaxMLC}
\end{equation}
It should be noted that from Eq.~(\ref{eq:AmaxSSC}), Eq.~(\ref{eq:AmaxDSC}), and  Eq.~(\ref{eq:AmaxMLC}), the maximum absorptance of SSPDs with single-side, double-side, and dielectric multi-layer optical cavities is described by the same formula.
From Eq.~(\ref{eq:dSSCmax}) and Eq.~(\ref{eq:dMLmax}), the thickness of the wire layer at the maximum absorptance is described by the same expression for both the single-side optical cavity and the dielectric multi-layer optical cavity.
In these cases, it depends on the refractive index of the input media.
In contrast, Eq.~(\ref{eq:dDSCmax}) indicates that, in the double-side optical cavity, the thickness of the wire layer at the maximum absorptance can be tuned through the combination of the refractive index of the input medium and the lower dielectric layer. 

\section*{Comparisons with numerical simulations}

\begin{table}[t]
\caption{Materials and refractive indices used for the analytical formulae and simulations. Note that the refractive index of the PEC mirror was set to $-1000i$ in the RCWA and FEM simulations, as it was not possible to set an infinite refractive index.}
  \label{tab:indecies}
  \centering
\begin{tabular}{ccc}
\hline
Material & Refractive index & Reference \\
\hline
Vacuum & 1 & \\
NbN & $4.905 - 4.293i$ &  \citeonline{yamashita2013low}\\
Si & 3.628 &   \citeonline{yamashita2013low}\\
SiO & 1.551 &   \citeonline{yamashita2013low}\\
SiO$_2$ & 1.444 &   \citeonline{yamashita2013low}\\
Ta$_2$O$_5$ & 2.15 &  \citeonline{zhang2017nbn}\\
PEC & $-1000i$ &  \\
Ag & $0.322 - 10.99i$ &   \citeonline{yamashita2013low}\\
\hline
\end{tabular}
\end{table}

\begin{figure*}[!t]
\centering\includegraphics[width=17cm]{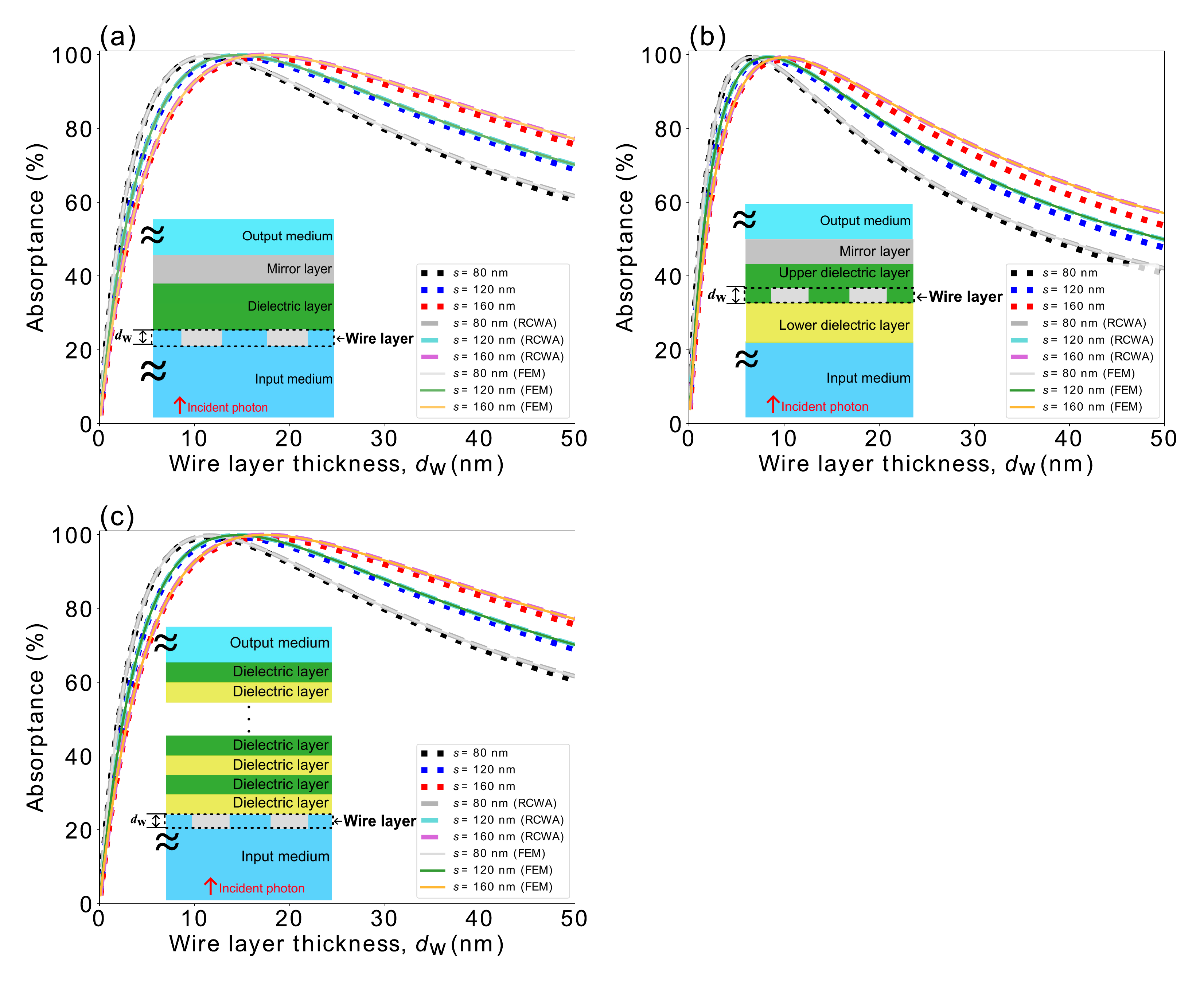}
\caption{Comparisons of analytical and simulation results. In each figure, the results represented by the dotted lines are calculated using the analytical formulae given by the transmission line model, and those represented by dashed~(solid) lines are obtained from the simulations using RCWA~(FEM). (a)~Dependence of the absorptance on the thickness of the wire layer of SSPDs with single-side optical cavities. (b)~ Dependence of the absorptance on the thickness of the wire layer of SSPDs with double-side optical cavities. (c)~Dependence of the absorptance on the thickness of the wire layer of SSPDs with dielectric multi-layer optical cavities.}
\label{fig:wire_calculation}
\end{figure*}

\begin{figure*}[!t]
\centering\includegraphics[width=17cm]{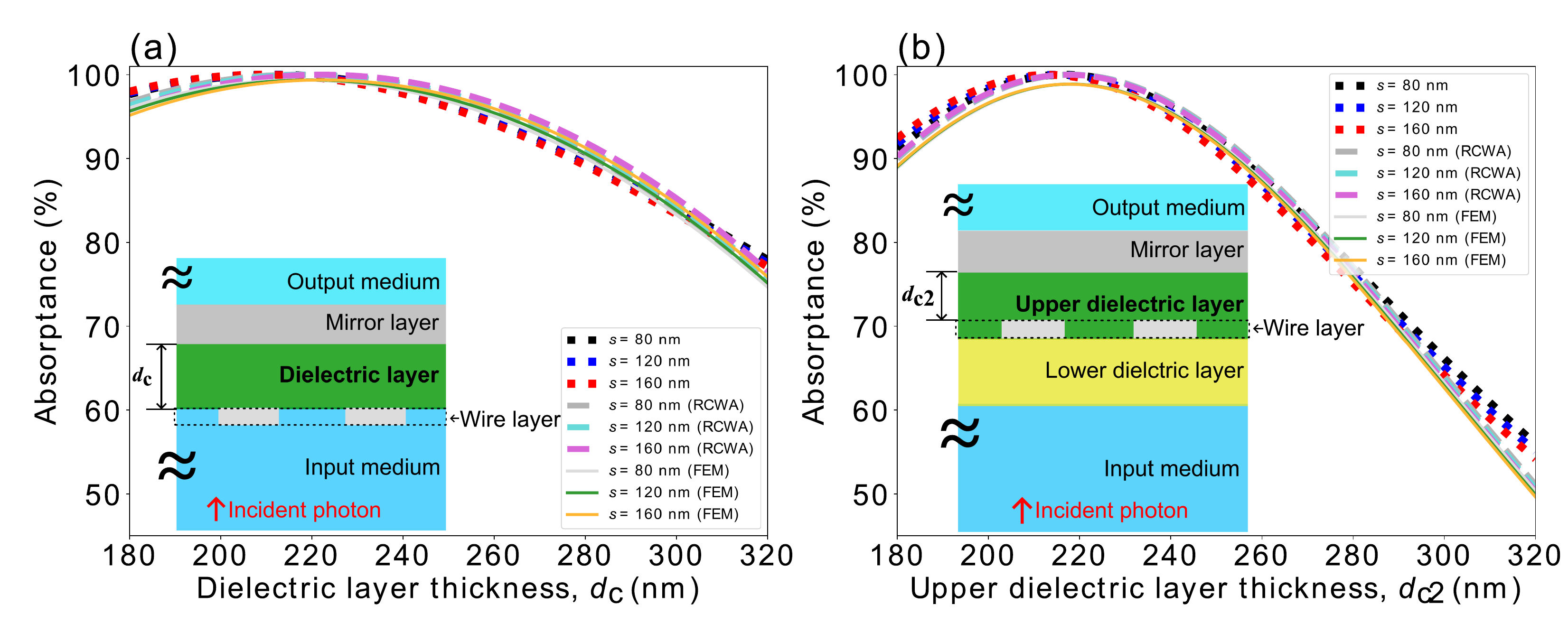}
\caption{Comparisons between analytical and simulation results. In each figure, the results represented by the dotted lines are calculated using the analytical formula given by the transmission line model, and those represented by the dashed~(solid) lines are obtained from simulations using RCWA~(FEM). (a)~Dependence of the absorptance on the thickness of the dielectric layer of SSPDs with single-side optical cavities. (b)~Dependence of the absorptance on the thickness of the upper dielectric layer of SSPDs with double-side optical cavities.}
\label{fig:dielectric_calculation}
\end{figure*}

Next, we compared the results obtained from the transmission line model with those from the simulation methods, RCWA and FEM. $S^4$~(Stanford Stratified Structure Solver)~\cite{liu2012s4} was used for the RCWA simulations, and COMSOL Multiphysics with RF module was used for the FEM simulations.
The derivation of the absorptance using COMSOL Multiphysics with RF module follows the method described in~\citeonline{anant2008optical}.
In these comparisons, the wavelength of the incident photon was set to $1550~{\rm nm}$. 
We used the refractive indices of the superconducting material, dielectric layer, metallic mirror in the optical cavities, and input medium reported in~\citeonline{yamashita2013low,zhang2017nbn} and summarized in Table~\ref{tab:indecies}.
Note that the refractive index of the PEC mirror was set to $-1000i$ in the RCWA and FEM simulations, as it was not possible to set an infinite refractive index.
In the comparisons, NbN was used as the superconducting wire, and the line width was fixed at $80~{\rm nm}$. 
The slit widths (filling factors) were set to $s = 80~{\rm nm}$ ($f = 0.50$), $s = 120~{\rm nm}$ ($f = 0.40$), and $s = 160~{\rm nm}$ ($f = 0.33$), satisfying Condition~I.
The thickness of the PEC or metallic mirror in SSPDs with single-side and double-side optical cavities in the simulation was fixed at $130~{\rm nm}$, which satisfied Condition~VI.
SSPDs with single-side and dielectric multi-layer optical cavities used vacuum as both the input and output media.
In SSPDs with single-side optical cavities, the dielectric layers were made of SiO, whereas SiO$_2$ and Ta$_2$O$_5$ were used in dielectric multi-layer optical cavities.
In the configuration of SSPDs with double-side optical cavities, Si was used as the input medium because the model assumes backside illumination through a Si substrate, which is commonly employed in practical device structure~\cite{miki2013high}, while the output medium was assumed to be vacuum.
In this case, when the Si substrate is sufficiently thick and an appropriate anti-reflection coating is formed between the vacuum and Si substrate, the incident light can be regarded as entering from Si, so that Si can be treated as the input medium. 
The lower and upper dielectric layers were assumed to be SiO$_2$ and SiO, respectively.

First, we compared the dependences of the absorptance on the thickness of the wire layer given by the analytical formulae for SSPDs with single-side optical cavities~(Eq.~(\ref{eq:ASSC})), double-side optical cavities~(Eq.~(\ref{eq:ADSC})), and dielectric multi-layer optical cavities~(Eq.~(\ref{eq:AML})) with the simulation results.
In these calculations and simulations, we fixed the thickness of the dielectric layer at one-quarter of the incident photon wavelength scaled by its refractive index.
Figure~\ref{fig:wire_calculation}~(a), (b), and (c) show the results for SSPDs with single-side, double-side, and dielectric multi-layer optical cavities, respectively.
In each figure,  the dotted lines show the results given by the analytical formulae, and dashed and solid lines indicate the RCWA and FEM simulation results, respectively. 
The analytical and simulation results are nearly identical.
Table~\ref{tab:cavity_comparisons} summarizes the thicknesses of the wire layer that achieve the maximum absorptance in SSPDs with single-side optical cavities~(Eq.~(\ref{eq:dSSCmax})), and double-side optical cavities~(Eq.~(\ref{eq:dDSCmax})), and dielectric multi-layer optical cavities~(Eq.~(\ref{eq:dMLmax})) and the corresponding results of RCWA and FEM.
The thickness of the wire layer that achieves the maximum absorptance differs by less than 2\% between the analytical results and the simulation results.
These results indicate that the analytical formulae can reliably determine the thickness of the wire layer required to achieve the maximum absorptance.
The slight difference between the analytical and  numerical results at larger wire thickness arises from the approximations introduced in Condition VII, which falls outside the range of validity in this regime.

Next, we compared the dependence of the absorptance on the thickness of the dielectric layer given by the analytical formula for SSPDs with single-side optical cavities~(Eq.~(\ref{eq:ASSC2})) and double-side optical cavities~(Eq.~(\ref{eq:ADSC2})) with that shown in the simulation results.
We assumed the material of the metallic mirror as Ag in both the optical cavities.
In the SSPDs with double-side optical cavities, the thickness of the lower dielectric layer was fixed at the quarter-wave thickness of the incident photon wavelength scaled by its refractive index. 
As shown in Fig.~\ref{fig:dielectric_calculation}, the analytical and simulation results are in good agreement. 
Table~\ref{tab:cavity_comparisons} summarizes the thickness of the dielectric layer that achieves the maximum absorptance in SSPDs with single-side optical cavities~(Eq.~(\ref{eq:dSSCcmax})) and double-side optical cavities~(Eq.~(\ref{eq:dc2max})).
The difference between the thickness of the dielectric layer to achieve maximum absorptance given by the analytical formulae and that given by simulations is less than 6\%.
Thus, the analytical and simulation results are nearly identical, confirming that the analytical approach is useful in obtaining the dielectric layer thickness at the maximum absorptance.
The slight discrepancy between analytical and simulation results increases as the thickness of dielectric layer deviates from quarter-wave thickness of the incident photon wavelength scaled by its refractive index, because the approximation introduced in Eq.~(\ref{eq:Fc}) falls outside its range of validity in this regime.
The source codes used to reproduce these results are provided in~\citeonline{source_codes}.

\begin{table}[t]
    \caption{Thickness of the wire layer or dielectric layer required to achieve the maximum absorptance given by the analytical formulae and by simulations using RCWA and FEM.}
    \label{tab:cavity_comparisons}
    \centering
    \begin{tabularx}{\textwidth}{X|XXX|X|XXX}
        \hline
        \multicolumn{8}{c}{\textbf{Single-side optical cavity}} \\ \hline
        \multicolumn{4}{c|}{$d^{\rm max}_{\rm w, SSC}$} & \multicolumn{4}{c}{$d^{\rm max}_{\rm c, SSC}$} \\ \hline
        Slit width & Eq.~(\ref{eq:dSSCmax}) & RCWA & FEM & Slit width  & Eq.~(\ref{eq:dSSCcmax}) & RCWA & FEM \\ \hline
        $80~{\rm \mu m}$ &  $11.6~{\rm nm}$ & $11.6~{\rm nm}$ & $11.6~{\rm nm}$ & $80~{\rm \mu m}$ & $211~{\rm nm}$ & $217~{\rm nm}$ & $217~{\rm nm}$ \\
        $120~{\rm \mu m}$ & $14.4~{\rm nm}$& $14.5~{\rm nm}$ & $14.5~{\rm nm}$ & $120~{\rm \mu m}$ & $210~{\rm nm}$ & $219~{\rm nm}$ & $219~{\rm nm}$ \\
        $160~{\rm \mu m}$ & $17.3~{\rm nm}$ & $17.4~{\rm nm}$ & $17.4~{\rm nm}$ & $160~{\rm \mu m}$ & $209~{\rm nm}$ & $222~{\rm nm}$ & $222~{\rm nm}$ \\
        \hline
        \multicolumn{8}{c}{\textbf{Double-side optical cavity}} \\ \hline
        \multicolumn{4}{c|}{$d^{\rm max}_{\rm w, DSC}$} & \multicolumn{4}{c}{$d^{\rm max}_{\rm c2, DSC}$} \\ \hline
        Slit width & Eq.~(\ref{eq:dDSCmax}) & RCWA & FEM & Slit width  & Eq.~(\ref{eq:dc2max}) & RCWA & FEM \\ \hline
        $80~{\rm \mu m}$ & $6.6~{\rm nm}$ & $6.6~{\rm nm}$ & $6.6~{\rm nm}$ & $80~{\rm \mu m}$ & $216~{\rm nm}$ & $218~{\rm nm}$ & $218~{\rm nm}$ \\
        $120~{\rm \mu m}$ & $8.2~{\rm nm}$ & $8.3~{\rm nm}$ & $8.3~{\rm nm}$ & $120~{\rm \mu m}$ & $215~{\rm nm}$ & $218~{\rm nm}$ & $218~{\rm nm}$ \\
        $160~{\rm \mu m}$ & $9.8~{\rm nm}$ & $10.0~{\rm nm}$ & $10.0~{\rm nm}$ & $160~{\rm \mu m}$ & $213~{\rm nm}$ & $217~{\rm nm}$ & $217~{\rm nm}$ \\
        \hline
        \multicolumn{8}{c}{\textbf{Dielectric multi-layer optical cavity}} \\ \hline
        \multicolumn{4}{c|}{$d^{\rm max}_{\rm w, MLC}$} & \multicolumn{4}{c}{} \\ \cline{1-4}
        Slit width & Eq.~(\ref{eq:dMLmax}) & RCWA & FEM & \multicolumn{4}{@{}c@{}}{} \\ \cline{1-4}
        $80~{\rm \mu m}$ & $11.6~{\rm nm}$ & $11.6~{\rm nm}$ & $11.6~{\rm nm}$ & \multicolumn{4}{@{}c@{}}{} \\ 
        $120~{\rm \mu m}$ & $14.4~{\rm nm}$ & $14.5~{\rm nm}$ & $14.5~{\rm nm}$ & \multicolumn{4}{@{}c@{}}{} \\ 
        $160~{\rm \mu m}$ & $17.3~{\rm nm}$ & $17.4~{\rm nm}$ & $17.4~{\rm nm}$ & \multicolumn{4}{@{}c@{}}{} \\ \hline
    \end{tabularx}
\end{table}

\section*{Discussion}

\subsection*{Physical mechanism to achieve maximum absorptance}

In this section, by introducing the impedance model, we clarify the physical mechanism by which the geometry of SSPDs with optical cavities achieves the maximum absorptance and show that it can be interpreted as an impedance matching condition between the SSPDs with optical cavity and the input medium.
The $F$ matrix is transformed into the input impedance $\eta_{\rm in}$, which is expressed as~\cite{pozar2011microwave}
\begin{equation}
    \eta_{\rm in} = \frac{F_{11}\eta_{\rm o} + F_{12}}{F_{21}\eta_{\rm o} + F_{22}}.
\label{eq:F2in}
\end{equation}
By substituting Eq.~(\ref{eq:FSSC}), Eq.~(\ref{eq:FDSC}), and Eq.~(\ref{eq:FML}) into Eq.~(\ref{eq:F2in}), the input impedances of SSPDs with single-side optical cavities~($\eta_{\rm in, SSC}$), double-side optical cavities~($\eta_{\rm in, DSC}$), and dielectric multi-layer optical cavities~($\eta_{\rm in, MLC}$) are expressed as
\begin{equation}
    \eta_{\rm in, SSC} = \frac{\eta_0}{ik_0\varepsilon_{\rm w}d_{\rm w}},
\label{eq:etainSSC}
\end{equation}
\begin{equation}
    \eta_{\rm in, DSC} = \frac{i\eta_0k_0\varepsilon_{\rm w}d_{\rm w}}{n_{\rm c1}^2},
\label{eq:etainDSC}
\end{equation}
and
\begin{equation}
    \eta_{\rm in, MLC} = \frac{\eta_0}{ik_0\varepsilon_{\rm w}d_{\rm w}}.
\label{eq:etainML}
\end{equation}
Here, we assume the PEC mirror, i.e., $\eta_{\rm o} = 0$, in SSPDs with single-side and double-side optical cavities.
For SSPDs with dielectric multi-layer optical cavities, the number of periods is sufficiently large.
By substituting Eq.~(\ref{eq:dSSCmax}) into Eq.~(\ref{eq:etainSSC}),  Eq.~(\ref{eq:dDSCmax}) into Eq.~(\ref{eq:etainDSC}), and Eq.~(\ref{eq:dMLmax}) into Eq.~(\ref{eq:etainML}), the input impedance of SSPDs with single-side optical cavities~($\eta^{\rm max}_{\rm in, SSC}$), double-side optical cavities~($\eta^{\rm max}_{\rm in, DSC}$), and dielectric multi-layer optical cavities~($\eta^{\rm max}_{\rm in, MLC}$), required to achieve the maximum absorptance, satisfies the following equations:
\begin{equation}
    \eta^{\rm max}_{\rm in, SSC} = \frac{\eta_{\rm i}|\varepsilon_{\rm w}|}{i\varepsilon_{\rm w}},
\end{equation}
\begin{equation}
    \eta^{\rm max}_{\rm in, DSC} = \frac{i\eta_{\rm i}\varepsilon_{\rm w}}{|\varepsilon_{\rm w}|},
\end{equation}
and
\begin{equation}
    \eta^{\rm max}_{\rm in, MLC} = \frac{\eta_{\rm i}|\varepsilon_{\rm w}|}{i\varepsilon_{\rm w}}.
\end{equation}
The results indicate that the absolute values of the input impedance for SSPDs with single-side, double-side, and dielectric multi-layer optical cavities should be identical to the input impedance, i.e., $|\eta_{\rm in}|=\eta_{\rm i}$, for achieving the maximum absorptance.
Therefore, the maximum absorptance is achieved when the impedance of the input medium matches the input impedance of SSPDs with optical cavities.
In other words, the reflectance becomes minimum.

\begin{figure}[!t]
\centering\includegraphics[width=8cm]{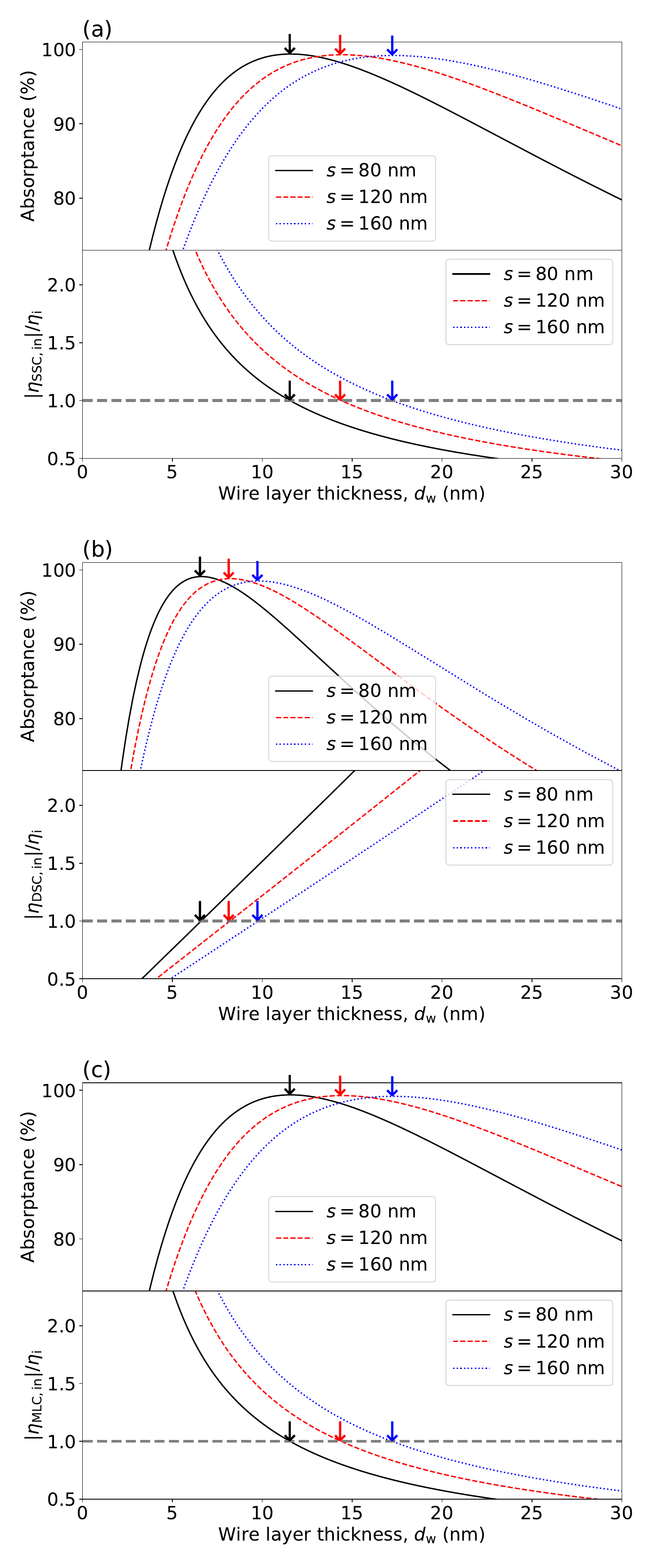}
\caption{Dependence of the absorptance and ratio of the input impedance to the impedance of the input media on the thickness of the wire layer. The upper panel of each figure shows the absorptance as a function of the thickness of the wire layer, determined by Eq.~(\ref{eq:ASSC}) in~(a), Eq.~(\ref{eq:ADSC}) in~(b), and Eq.~(\ref{eq:AML}) in~(c). The lower panel shows the ratio of the absolute value of the input impedance of SSPDs with the cavities to the impedance of the input medium. (a) SSPDs with single-side optical cavities.~(b) SSPDs with double-side optical cavities.~(c) SSPDs with dielectric multi-layer optical cavities. The arrows indicate the maximum absorptance and corresponding ratio of the input impedance to the impedance of the input media.}
\label{fig:impedance}
\end{figure}

Figure~\ref{fig:impedance} shows the dependence of the absorptance and ratio of input impedance to impedance of the input medium on the thickness of SSPDs with optical cavities.
In this calculation, the geometry of SSPDs with optical cavities is the same as that used in the previous section.
At the maximum absorptance, the absolute value of the input impedance of SSPDs with optical cavities $|\eta_{\rm in}|$ approaches the impedance of the input media $\eta_{\rm i}$, i.e., impedance matching is achieved.
The source codes used to reproduce these results are provided in~\citeonline{source_codes}.

\subsection*{Physical role of lower dielectric layer in double-side optical cavity}

We investigated the physical role of the lower dielectric layer by applying the quarter-wave impedance transformer~\cite{pozar2011microwave} to SSPDs with double-side optical cavities.
In SSPDs with double-side optical cavities, the input impedance from the lower dielectric layer, $\eta'_{\rm in, DSC}$, is given by
\begin{equation}
    \eta'_{\rm in, DSC} = \frac{\eta_0}{ik_0\varepsilon_{\rm w}d_{\rm w}}
\label{eq:etadDSC}
\end{equation}
in the case where the metallic mirror is assumed to be a PEC mirror.
The quarter-wave transformer has the characteristic impedance that satisfies $\eta_{\rm QWT} = \sqrt{\eta_{\rm i}\eta_{\rm o}}$, where $\eta_{\rm i}$ ($\eta_{\rm o}$) represents the impedance of the input (output) medium.
In SSPDs with double-side optical cavities, by replacing $\eta_{\rm o}$ with $\eta'_{\rm in, DSC}$, a $\eta_{\rm QWT}$ is given by
\begin{equation}
    \eta_{\rm QWT} = \sqrt{\frac{\eta_0\eta_{\rm i}}{k_0\varepsilon_{\rm w}d_{\rm w}}}.
\label{eq:etaQWT}
\end{equation}
From Eq.~(\ref{eq:etaQWT}), we can derive
\begin{equation}
    d_{\rm w} = \frac{n_{\rm QWT}^2}{k_0n_{\rm i}|\varepsilon_{\rm w}|}.
\label{eq:dwQWT}
\end{equation}
The result indicates that the lower dielectric layer in the double-side optical cavities works as a quarter-wave impedance transformer, as Eq.~(\ref{eq:dwQWT}) is identical to Eq.~(\ref{eq:dDSCmax}) when $n_{\rm QWT}\rightarrow n_{\rm c1}$.
Therefore, the absorptance of SSPDs with double-side optical cavities depends on the refractive index of the lower dielectric layer.
In other words, the refractive index of the lower dielectric layer needs to be equal to Eq.~(\ref{eq:etaQWT}), i.e., $\eta_{\rm c1} = \eta_{\rm QWT}$, to achieve the maximum absorptance.

\section*{Device Design Guidelines for SSPDs with Optical Cavities}

In this section, we summarize a practical design flow for SSPDs with optical cavities based on the analytical results obtained from transmission line and impedance models.
The first step is to define the target wavelength at which high absorptance is required. 
Next, the cavity type, superconducting material, dielectric material, mirror material, and filling factor are selected according to the application requirements and fabrication constraints.
In general, different cavity types involve different design characteristics and trade-offs.
For example, the single-side optical cavity has a simple structure and is easy to fabricate, whereas the double-side optical cavity provides additional design flexibility through the combination of the input medium and the lower dielectric layer.
The dielectric multi-layer optical cavity can achieve high absorptance over a broad bandwidth for suitable combinations of the two dielectric refractive indices, although it requires a more complex multi-layer fabrication process.
Note that, in SSPDs with double-side optical cavities, the refractive index of the lower dielectric layer should be chosen to satisfy the role of quarter-wave impedance transformer to achieve the maximum absorptance. 
In SSPDs with dielectric multi-layer optical cavities, the layers are stacked such that the dielectric layer adjacent to the superconducting wire has a smaller refractive index than the next layer, i.e. $n_{\rm c1} < n_{\rm c2}$ in Fig.~\ref{fig:photon_cavity_type}~(c).
The thickness of wire layer are chosen to satisfy Eq.~(\ref{eq:dSSCmax}) in SSPDs with single-side optical cavities, Eq.~(\ref{eq:dDSCmax}) in SSPDs with double-side optical cavities, and Eq.~(\ref{eq:dMLmax}) in SSPDs with dielectric multi-layer optical cavities.
Similarly, the thickness of dielectric layer are chosen to satisfy Eq.~(\ref{eq:dSSCcmax}) in SSPDs with single-side optical cavities, and Eq.~(\ref{eq:dphiDSC}) in SSPDs with double-side optical cavities.
Finally, using the impedance model given by Eq.~(\ref{eq:F2in}), one can check that the input impedance of the SSPDs with optical cavities ($\eta_{\rm in}$) is close to 
the impedance of the input medium ($\eta_{\rm i}$), which ensures the maximum absorptance.

\section*{Conclusions}

We derived the analytical formulae for the absorptance of SSPDs with optical cavities, the maximum absorptance, and required thickness of the wire layer and dielectric layer to achieve the maximum absorptance by applying the transmission line model. 
We compared the results given by the analytical formulae proposed in this study with those obtained from simulations conducted using RCWA and FEM.
These comparisons showed that the results are nearly identical.
We investigated the physical mechanism underlying the maximum absorptance of SSPDs with optical cavities by introducing the impedance model.
The maximum absorptance of SSPDs with optical cavities is achieved when the impedance of the input medium and input impedance of SSPDs with optical cavities match.
We clarified that the lower dielectric layer in the double-side optical cavities works as a quarter-wave impedance transformer, and reported the refractive index of the lower dielectric layer required to achieve the maximum absorptance.
The design concepts proposed in this study are applicable to other superconducting detectors with optical cavities, such as MKIDs~\cite{day2003broadband} and transition-edge sensors (TESs)~\cite{irwin1995application}.

%\subsection*{Subsection}

%Example text under a subsection. Bulleted lists may be used where appropriate, e.g.

%\begin{itemize}
%\item First item
%\item Second item
%\end{itemize}

%\subsubsection*{Third-level section}
 
%Topical subheadings are allowed.

%\section*{Discussion}

%The Discussion should be succinct and must not contain subheadings.

%\section*{Methods}

%Topical subheadings are allowed. Authors must ensure that their Methods section includes adequate experimental and characterization data necessary for others in the field to reproduce their work

\section*{Data Availability}
The source code to reproduce the figures and table is available in Zenodo with the identifier DOI: \url{https://doi.org/10.5281/zenodo.19722815}. 

\section*{Funding}

This work was supported in part by JST PRESTO grant No. JPMJPR23F4 and JSPS KAKENHI grant No. 23K17323. 

\bibliography{sample}

@article{Goltsman:01,
author = {G. N. Gol'tsman and O. Okunev and G. Chulkova and A. Lipatov and A. Semenov and K. Smirnov and B. Voronov and A. Dzardanov and C. Williams and Roman Sobolewski},
title = {{Picosecond superconducting single-photon optical detector}},
journal = {Applied Physics Letters},
number = {6},
pages = {705--707},
publisher = {AIP Publishing},
volume = {79},
month = {August},
year = {2001},
doi = {10.1063/1.1388868},
}

@article{zhong2020quantum,
title={{Quantum computational advantage using photons}},
author={Han-Sen Zhong and Hui Wang and Yu-Hao Deng and Ming-Cheng Chen and Li-Chao Peng and Yi-Han Luo and Jian Qin and Dian Wu and Xing Ding and Yi Hu, Peng Hu and Xiao-Yan Yang and Wei-Jun Zhang and Hao Li and Yuxuan Li and Xiao Jiang and Lin Gan and Guangwen Yang and Lixing You and Zhen Wang and Li Li and Nai-Le Liu and Chao-Yang Lu and Jian-Wei Pan},
journal = {Science},
number = {6523},
pages = {1460--1463},
publisher={American Association for the Advancement of Science},
volume = {370},
month = {December},
year={2020},
doi = {10.1126/science.abe8770}
}

@article{sasaki2011field,
title={{Field test of quantum key distribution in the Tokyo QKD Network}},
author={M. Sasaki and M. Fujiwara and H. Ishizuka and W. Klaus and K. Wakui and M. Takeoka and S. Miki and T. Yamashita and Z. Wang and A. Tanaka and K. Yoshino and Y. Nambu and S. Takahashi and  A. Tajima and A. Tomita and T. Domeki and T. Hasegawa and Y. Sakai and H. Kobayashi and T. Asai and K. Shimizu and T. Tokura and T. Tsurumaru and M. Matsui and T. Honjo and K. Tamaki and H. Takesue and Y. Tokura and J. F. Dynes and A. R. Dixon and A. W. Sharpe and Z. L. Yuan and A. J. Shields and S. Uchikoga and M. Legr\'{e} and S. Robyr and P. Trinkler and L. Monat and  J.-B. Page and G. Ribordy and A. Poppe and A. Allacher and O. Maurhart and T. L\"{a}nger and  M. Peev and A. Zeilinger},
journal={Optics Express},
number={11},
pages={10387--10409},
publisher={Optica Publishing Group},
volume={19},
month = {May},
year={2011},
doi = {10.1364/OE.19.010387}
}

@article{kobayashi2016frequency,
title = {{Frequency-domain Hong–Ou–Mandel interference}},
author={Toshiki Kobayashi and Rikizo Ikuta and Shuto Yasui and Shigehito Miki and Taro Yamashita and Hirotaka Terai and Takashi Yamamoto and Masato Koashi and Nobuyuki Imoto},
journal = {Nature Photonics},
number = {7},
pages = {441--444},
publisher = {Nature Publishing Group UK London},
volume={10},
month = {April},
year = {2016},
doi = {https://doi.org/10.1038/nphoton.2016.74}
}

@article{endo2021quantum,
title = {{Quantum detector tomography of a superconducting nanostrip photon-number-resolving detector}},
author={Mamoru Endo and Tatsuki Sonoyama and Mikihisa Matsuyama and Fumiya Okamoto and Shigehito Miki and Masahiro Yabuno and Fumihiro China and Hirotaka Terai and Akira Furusawa},
journal = {Optics Express},
number = {8},
pages = {11728--11738},
publisher = {Optica Publishing Group},
volume = {29},
month = {March},
year = {2021},
doi = {10.1364/OE.423142}
}

@article{marsili2013detecting,
title={{Detecting single infrared photons with 93\% system efficiency}},
author={F. Marsili and V. B. Verma and J. A. Stern and S. Harrington and A. E. Lita and T. Gerrits and I. Vayshenker and B. Baek and M. D. Shaw and R. P. Mirin and S. W. Nam},
journal={Nature Photonics},
number={3},
pages={210--214},
publisher={Nature Publishing Group UK London},
volume={7},
month = {March},
year={2013},
doi = {10.1038/nphoton.2013.13}
}

@article{zhang2017nbn,
title={{NbN superconducting nanowire single photon detector with efficiency over 90\% at 1550 nm wavelength operational at compact cryocooler temperature}},
author={WeiJun Zhang and LiXing You and Hao Li and Jia Huang and ChaoLin Lv and Lu Zhang and XiaoYu Liu and JunJie Wu and Zhen Wang and XiaoMing Xie},
journal={Science China Physics, Mechanics \& Astronomy},
pages={1--10},
publisher={Springer},
volume={60},
month={October},
year={2017},
doi={10.1007/s11433-017-9113-4}
}

@article{reddy2020superconducting,
title={{Superconducting nanowire single-photon detectors with 98\% system detection efficiency at 1550 nm}},
author={Dileep V. Reddy and Robert R. Nerem and Sae Woo Nam and Richard P. Mirin and Varun B. Verma},
journal={Optica},
number={12},
pages={1649--1653},
publisher={Optica Publishing Group},
volume={7},
month={December},
year={2020},
doi={/10.1364/OPTICA.400751}
}

@article{hu2020detecting,
title={{Detecting single infrared photons toward optimal system detection efficiency}},
author={Peng Hu and Hao Li and Lixing You and Heqing Wang and You Xiao and Jia Huang and Xiaoyan Yang and Weijun Zhang and Zhen Wang and Xiaoming Xie},
journal={Optics Express},
number={24},
pages={36884--36891},
publisher={Optica Publishing Group},
volume={28},
year={2020},
doi={10.1364/OE.410025}
}

@article{chang2021detecting,
title={{Detecting telecom single photons with $(99.5_{-2.07}^{+0.5})$\% system detection efficiency and high time resolution}},
author={J. Chang and J. W. N. Los and J. O. Tenorio-Pearl and N. Noordzij and R. Gourgues and A. Guardiani and J. R. Zichi and S. F. Pereira and H. P. Urbach and V. Zwiller and S. N. Dorenbos and I. Esmaeil Zadeh},
journal={APL Photonics},
number={3},
pages={036114},
publisher={AIP Publishing},
volume={6},
month={March},
year={2021},
doi={10.1063/5.0039772}
}

@article{xu2021superconducting,
title={{Superconducting microstrip single-photon detector with system detection efficiency over 90\% at 1550 nm}},
author={Guang-Zhao Xu and Wei-Jun Zhang and Li-Xing You and Jia-Min Xiong and Xing-Qu Sun and Hao Huang and Xin Ou and Yi-Ming Pan and Chao-Lin Lv and Hao Li and Zhen Wang and Xiao-Ming Xie
},
journal={Photonics Research},
number={6},
pages={958--967},
publisher={Optica Publishing Group},
volume={9},
month={June},
year={2021},
doi={10.1364/PRJ.419514}
}

@article{reddy2022broadband,
title={{Broadband polarization insensitivity and high detection efficiency in high-fill-factor superconducting microwire single-photon detectors}},
author={Dileep V. Reddy and Negar Otrooshi and Sae Woo Nam and Richard P. Mirin and Varun B. Verma},
journal={APL Photonics},
number={5},
pages={051302},
publisher={AIP Publishing},
volume={7},
month={May},
year={2022},
doi={10.1063/5.0088007}
}

@article{miki2018superconducting,
title={{Superconducting coincidence photon detector with short timing jitter}},
author={S. Miki and S. Miyajima and M. Yabuno and T. Yamashita and T. Yamamoto and N. Imoto and R. Ikuta and R. A. Kirkwood and R. H. Hadfield and H. Terai},
number={26},
pages={262601},
publisher={AIP Publishing},
journal={Applied Physics Letters},
volume={112},
year={2018},
doi={10.1063/1.5037254}
}

@article{korzh2020demonstration,
title={{Demonstration of sub-3 ps temporal resolution with a superconducting nanowire single-photon detector}},
author={Boris Korzh and Qing-Yuan Zhao and Jason P. Allmaras and Simone Frasca and Travis M. Autry and Eric A. Bersin and Andrew D. Beyer and Ryan M. Briggs and Bruce Bumble and Marco Colangelo and Garrison M. Crouch and Andrew E. Dane and Thomas Gerrits and Adriana E. Lita and Francesco Marsili and Galan Moody and Cristi\'{a}n Pe\~{n}a and Edward Ram\'{i}rez and Jake D. Rezac and Neil Sinclair and Martin J. Stevens and \'{A}ngel E. Velasco and Varun B. Verma and Emma E. Wollman and Si Xie and Di Zhu and Paul D. Hale and Mar\'{i}a Spiropulu and Kevin L. Silverman and Richard P. Mirin and Sae Woo Nam and Alexander G. Kozorezov and Matthew D. Shaw and Karl K. Berggren
},
journal={Nature Photonics},
number={4},
pages={250--255},
publisher={Nature Publishing Group UK London},
volume={14},
month={March},
year={2020},
doi={10.1038/s41566-020-0589-x}
}

@article{colangelo2023impedance,
title={{Impedance-matched differential superconducting nanowire detectors}},
author={Marco Colangelo and Boris Korzh and Jason P. Allmaras and Andrew D. Beyer and Andrew S. Mueller and Ryan M. Briggs and Bruce Bumble and Marcus Runyan and Martin J. Stevens and Adam N. McCaughan and Di Zhu and Stephen Smith and Wolfgang Becker and Lautaro Narváez and Joshua C. Bienfang and Simone Frasca and Angel E. Velasco and Edward E. Ramirez and Alexander B. Walter and Ekkehart Schmidt and Emma E. Wollman and Maria Spiropulu and Richard Mirin and Sae Woo Nam and Karl K. Berggren and Matthew D. Shaw},
journal={Physical Review Applied},
number={4},
publisher={APS},
pages={044093},
volume={19},
month={April},
year={2023},
doi={10.1103/PhysRevApplied.19.044093}
}

@article{yang2014superconducting,
title={{Superconducting nanowire single photon detector with on-chip bandpass filter}},
author={Xiaoyan Yang and Hao Li and Weijun Zhang and Lixing You and Lu Zhang and Xiaoyu Liu and Zhen Wang and Wei Peng and Xiaoming Xie and Mianheng Jiang},
journal={Optics Express},
number={13},
pages={16267--16272},
publisher={Optica Publishing Group},
volume={22},
month = {April},
year={2014},
doi={https://doi.org/10.1364/OE.22.016267}
}

@article{shibata2015ultimate,
title={{Ultimate low system dark-count rate for superconducting nanowire single-photon detector}},
author={Hiroyuki Shibata and Kaoru Shimizu and Hiroki Takesue and Yasuhiro Tokura},
journal={Optics Letters},
number={14},
pages={3428--3431},
publisher={Optica Publishing Group},
volume={40},
month={July},
year={2015},
doi={10.1364/OL.40.003428}
}

@article{zhang2018fiber,
title={{Fiber-coupled superconducting nanowire single-photon detectors integrated with a bandpass filter on the fiber end-face}},
author={Zhang, WJ and Yang, XY and Li, H and You, LX and Lv, CL and Zhang, L and Zhang, CJ and Liu, XY and Wang, Z and Xie, XM},
journal={Superconductor Science and Technology},
number={3},
pages={035012},
publisher={IOP Publishing},
volume={31},
month={February},
year={2018},
doi={10.1088/1361-6668/aaa6b4}
}

@article{chiles2022new,
title={{New constraints on dark photon dark matter with superconducting nanowire detectors in an optical haloscope}},
author={Jeff Chiles and Ilya Charaev and Robert Lasenby and Masha Baryakhtar and Junwu Huang and Alexana Roshko and George Burton and Marco Colangelo and Ken Van Tilburg and Asimina Arvanitaki and Sae Woo Nam and Karl K. Berggren},
journal={Physical Review Letters},
number={23},
pages={231802},
publisher={APS},
volume={128},
month={June},
year={2022},
doi={10.1103/PhysRevLett.128.231802}
}

@article{zhang201916,
title={{A 16-pixel interleaved superconducting nanowire single-photon detector array with a maximum count rate exceeding 1.5~GHz}},
author={Weijun Zhang and Jia Huang and Chengjun Zhang and Lixing You and Chaolin Lv and Lu Zhang and Hao Li and Zhen Wang and Xiaoming Xie},
journal={IEEE Trans. Appl. Supercond.},
number={5},
pages={1--4},
publisher={IEEE},
volume={29},
month={January},
year={2019},
doi={10.1109/TASC.2019.2895621}
}

@article{resta2023gigahertz,
title={{Gigahertz detection rates and dynamic photon-number resolution with superconducting nanowire arrays}},
author={Giovanni V. Resta and Lorenzo Stasi and Matthieu Perrenoud and Sylvain El-Khoury and Tiff Brydges and Rob Thew and Hugo Zbinden and F\'{e}lix Bussi\`{e}res},
journal={Nano Letters},
number={13},
pages={6018--6026},
publisher={ACS Publications},
volume={23},
year={2023},
month={June},
doi={10.1021/acs.nanolett.3c01228.}
}

@article{craiciu2023high,
title={{High-speed detection of 1550 nm single photons with superconducting nanowire detectors}},
author={Ioana Craiciu and Boris Korzh and Andrew D. Beyer and Andrew Mueller and Jason P. Allmaras and Lautaro Narváez and Maria Spiropulu and Bruce Bumble and Thomas Lehner and Emma E. Wollman and Matthew D. Shaw},
journal={Optica},
number={2},
pages={183--190},
publisher={Optica Publishing Group},
volume={10},
month={January},
year={2023},
doi={10.1364/OPTICA.478960}
}

@article{hao2024compact,
title={{A compact multi-pixel superconducting nanowire single-photon detector array supporting gigabit space-to-ground communications}},
author={Hao Hao and Qing-Yuan Zhao and Yang-Hui Huang and Jie Deng and Fan Yang and Sai-Ying Ru and Zhen Liu and Chao Wan and Hao Liu and Zhi-Jian Li and Hua-Bing Wang and Xue-Cou Tu and La-Bao Zhang and Xiao-Qing Jia and Xing-Long Wu and Jian Chen and Lin Kang and Pei-Heng Wu},
journal={Light: Science Applications},
number={1},
pages={25},
publisher={Nature Publishing Group UK London},
volume={13},
month={January},
year={2024},
doi={10.1038/s41377-023-01374-1}
}

@article{verevkin2002detection,
title={{Detection efficiency of large-active-area NbN single-photon superconducting detectors in the ultraviolet to near-infrared range}},
author={A. Verevkin and J. Zhang and Roman Sobolewski and A. Lipatov and O. Okunev and G. Chulkova and A. Korneev and K. Smirnov and G. N. Gol\'{t}sman and A. Semenov},
journal={Applied Physics Letters},
number={25},
pages={4687--4689},
publisher={AIP Publishing},
volume={80},
month={June},
year={2002},
doi={10.1063/1.1487924}
}

@article{rosfjord2006nanowire,
title={{Nanowire single-photon detector with an integrated optical cavity and anti-reflection coating}},
author={Kristine M. Rosfjord and Joel K. W. Yang and Eric A. Dauler and Andrew J. Kerman and Vikas Anant and Boris M. Voronov and Gregory N. Gol\'{t}sman and Karl K. Berggren},
journal={Optics Express},
number={2},
pages={527--534},
publisher={Optica Publishing Group},
volume={14},
month={January},
year={2006},
doi={10.1364/OPEX.14.000527}
}

@article{kerman2007constriction,
title={{Constriction-limited detection efficiency of superconducting nanowire single-photon detectors}},
author={Andrew J. Kerman and Eric A. Dauler and Joel K. W. Yang and Kristine M. Rosfjord and Vikas Anant and Karl K. Berggren and Gregory N. Gol\'{t}sman and Boris M. Voronov},
journal={Applied Physics Letters},
number={10},
pages={101110},
publisher={AIP Publishing},
volume={90},
month={March},
year={2007},
doi={10.1063/1.2696926}
}

@article{miki2009compactly,
title={{Compactly packaged superconducting nanowire single-photon detector with an optical cavity for multichannel system}},
author={Shigehito Miki and Masanori Takeda and Mikio Fujiwara and Masahide Sasaki and Zhen Wang},
journal={Optics Express},
number={26},
pages={23557--23564},
publisher={Optica Publishing Group},
volume={17},
month={December},
year={2009},
doi={10.1364/OE.17.023557}
}

@article{miki2010multichannel,
title={{Multichannel SNSPD system with high detection efficiency at telecommunication wavelength}},
author={Shigehito Miki and Taro Yamashita and Mikio Fujiwara and Masahide Sasakiand and Zhen Wang},
journal={Optics Letters},
number={13},
pages={2133--2135},
publisher={Optica Publishing Group},
volume={35},
month={July},
year={2010},
doi={https://doi.org/10.1364/OL.35.002133}
}

@article{gaggero2010nanowire,
title={{Nanowire superconducting single-photon detectors on {GaAs} for integrated quantum photonic applications}},
author={A. Gaggero and S. Jahanmiri Nejad and F. Marsili and F. Mattioli and R. Leoni and D. Bitauld and D. Sahin and G. J. Hamhuis and R. N\"{o}tzel and R. Sanjines and A. Fiore},
journal={Applied Physics Letters},
number={15},
pages={151108},
volume={97},
month={Octorber},
year={2010},
publisher={AIP Publishing},
doi={10.1063/1.3496457}
}

@article{miller2011compact,
title={{Compact cryogenic self-aligning fiber-to-detector coupling with losses below one percent}},
author={Aaron J. Miller and Adriana E. Lita and Brice Calkins and Igor Vayshenker and Steven M. Gruber and Sae Woo Nam
},
journal={Optics Express},
number={10},
publisher={Optica Publishing Group},
volume={19},
pages={9102--9110},
month={March},
year={2011},
doi={10.1364/OE.19.009102}
}

@article{clem2011geometry,
title={{Geometry-dependent critical currents in superconducting nanocircuits}},
author={John R. Clem and Karl K. Berggren},
number={17},
journal={Physical Review B},
publisher={APS},
volume={84},
pages={174510},
month={November},
year={2011},
doi={10.1103/PhysRevB.84.174510}
}

@article{miki2013high,
title={{High performance fiber-coupled NbTiN superconducting nanowire single photon detectors with Gifford-McMahon cryocooler}},
author={Shigehito Miki and Taro Yamashita and Hirotaka Terai and Zhen Wang},
journal={Optics Express},
number={8},
publisher={Optica Publishing Group},
volume={21},
pages={10208--10214},
month={April},
year={2013},
doi={10.1364/OE.21.010208}
}

@article{fang2020ingaas,
title={{InGaAs/InP single-photon detectors with 60\% detection efficiency at 1550 nm}},
author={Yu-Qiang Fang and Wei Chen and Tian-Hong Ao and Cong Liu and Li Wang and Xin-Jiang Gao and Jun Zhang and Jian-Wei Pan},
journal={Rev. Sci. Instrum.},
number={8},
publisher={AIP Publishing},
volume={91},
month={August},
year={2020},
doi={10.1063/5.0014123}
}

@article{kamada2008efficient,
title={{Efficient and low-noise single-photon detection in 1550 nm communication band by frequency upconversion in periodically poled LiNbO$_3$ waveguides}},
author={H. Kamada and M. Asobe and T. Honjo and H. Takesue and Y. Tokura, Y. Nishida and O. Tadanaga and and H. Miyazawa},
journal={Optics Letters},
number={7},
publisher={Optica Publishing Group},
volume={33},
pages={639--641},
month={March},
year={2008},
doi={10.1364/OL.33.000639}
}

@article{anant2008optical,
title={{Optical properties of superconducting nanowire single-photon detectors}},
author={Vikas Anant and Andrew J. Kerman and Eric A. Dauler and Joel K. W. Yang and Kristine M. Rosfjord and Karl K. Berggren},
journal={Optics Express},
number={14},
publisher={Optica Publishing Group},
volume={16},
pages={10750--10761},
month={July},
year={2008},
doi={10.1364/OE.16.010750}
}

@article{yamashita2013low,
title={{Low-filling-factor superconducting single photon detector with high system detection efficiency}},
author={Taro Yamashita and Shigehito Miki and Hirotaka Terai and Zhen Wang},
journal={Optics Express},
number={22},
publisher={Optica Publishing Group},
volume={21},
pages={27177--27184},
month={November},
year={2013},
doi={10.1364/OE.21.027177}
}

@article{moharam1981rigorous,
title={{Rigorous coupled-wave analysis of planar-grating diffraction}},
author={M. G. Moharam and T. K. Gaylord},
journal={JOSA},
number={7},
publisher={Optica Publishing Group},
volume={71},
pages={811--818},
month={July},
year={1981},
doi={10.1364/JOSA.71.000811}
}

@article{turner1956stiffness,
title={{Stiffness and deflection analysis of complex structures}},
author={M. J. Turner and R. W. Clough and H. C. Martin and L. J. Topp},
journal={Journal of the Aeronautical Sciences},
number={9},
volume={23},
pages={805--823},
year={1956},

}

@book{pozar2011microwave,
  title={{Microwave engineering: theory and techniques, 4th Edition}},
  author={Pozar, David M},
  year={2011},
  publisher={John Wiley \& Sons}
}

@article{driessen2009impedance,
title={{Impedance model for the polarization-dependent optical absorption of superconducting single-photon detectors}},
author={E. F. C. Driessen and F. R. Braakman and E. M. Reiger and S. N. Dorenbos and V. Zwiller and M. J. A. de Dood},
journal={Eur. Phys. J. - Appl. Phys.},
number={1},
publisher={EDP sciences},
volume={47},
pages={10701},
month={May},
year={2009},
doi={10.1051/epjap/2009087}
}

@article{kouwenhoven2022model,
title={{Model and measurements of an optical stack for broadband visible to near-infrared absorption in TiN MKIDs}},
author={K. Kouwenhoven and I. Elwakil and J. van Wingerden and V. Murugesan and D. J. Thoen and J. J. A. Baselmans  and P. J. de Visser},
journal={Journal of Low Temperature Physics},
number={5},
publisher={Springer},
volume={209},
pages={1249--1257},
month={July},
year={2022},
doi={10.1007/s10909-022-02774-0}
}

@article{day2003broadband,
title={{A broadband superconducting detector suitable for use in large arrays}},
author={Peter K. Day and Henry G. LeDuc and Benjamin A. Mazin and Anastasios Vayonakis and Jonas Zmuidzinas },
journal={Nature},
number={6960},
publisher={Nature Publishing Group UK London},
volume={425},
pages={817--821},
month={October},
year={2003},
doi={10.1038/nature02037}
}

@article{yabuno2023superconducting,
title={{Superconducting wide strip photon detector with high critical current bank structure}},
author={Masahiro Yabuno and Fumihiro China and Hirotaka Terai and Shigehito Miki},
journal={Optica Quantum},
number={1},
publisher={Optica Publishing Group},
volume={1},
pages={26--34},
month={October},
year={2023},
doi={10.1364/OPTICAQ.497675}
}

@article{frickey1994conversions,
title={{Conversions between S, Z, Y, H, ABCD, and T parameters which are valid for complex source and load impedances}},
author={D.A. Frickey},
journal={IEEE Trans. Microw. Theory Tech.},
number={2},
publisher={IEEE},
volume={42},
pages={205--211},
month={February},
year={1994},
doi={10.1109/22.275248}
}

@article{aspnes1982local,
title={{Local-field effects and effective-medium theory: {A} microscopic perspective}},
author={D. E. Aspnes},
journal={Am. J. Phys},
number={8},
volume={50},
pages={704--709},
month={September},
year={1982}
}

@article{irwin1995application,
title={{An application of electrothermal feedback for high resolution cryogenic particle detection}},
author={K. D. Irwin},
journal={Applied Physics Letters},
number={15},
volume={66},
publisher={American Institute of Physics},
pages={1998--2000},
month={April},
year={1995},
doi={10.1063/1.113674}
}

@article{liu2012s4,
title={{S4: A free electromagnetic solver for layered periodic structures}},
author={Victor Liu and Shanhui Fan},
journal={Computer Physics Communications},
number={10},
volume={183},
publisher={Elsevier},
pages={2233--2244},
month={October},
year={2012},
doi={10.1016/j.cpc.2012.04.026}
}

@misc{source_codes,
  author  = {H. Kutsuma and T. Yamashita},
  title   = {Source codes for 'Elucidating mechanism of optical cavities in superconducting strip single photon detectors using transmission line and impedance models'},
  year    = {2025},
  publisher = {Zenodo},
  doi     = {https://doi.org/10.5281/zenodo.19722815},
}

%\noindent LaTeX formats citations and references automatically using the bibliography records in your .bib file, which you can edit via the project menu. Use the cite command for an inline citation, e.g.  \cite{Hao:gidmaps:2014}.

%For data citations of datasets uploaded to e.g. \emph{figshare}, please use the \verb|howpublished| option in the bib entry to specify the platform and the link, as in the \verb|Hao:gidmaps:2014| example in the sample bibliography file.

\section*{Acknowledgements}

Hiroki Kutsuma and Taro Yamashita acknowledge the Center for Heterogeneous Quantum/Material Fusion Technologies and Center for Key Interdisciplinary Research at Tohoku University.

\section*{Author contributions statement}

%Must include all authors, identified by initials, for example:
%A.A. conceived the experiment(s),  A.A. and B.A. conducted the experiment(s), C.A. and D.A. analysed the results.  All authors reviewed the manuscript. 

H.K. conceived the project, developed the software, performed simulations, and wrote the manuscript.
T.Y. supervised the project, contributed to the discussion of the study, and the revision of the manuscript.

\section*{Additional information}

\textbf{Accession codes}: The source code to reproduce the figures and table is available in Zenodo with the identifier DOI: \url{https://doi.org/10.5281/zenodo.19722815}.\textbf{Competing interests}: The authors declare no competing interests.

%The corresponding author is responsible for submitting a \href{http://www.nature.com/srep/policies/index.html#competing}{competing interests statement} on behalf of all authors of the paper. This statement must be included in the submitted article file.

%\begin{figure}[hdtp]
%\centering
%\includegraphics[width=\linewidth]{stream}
%\caption{Legend (350 words max). Example legend text.}
%\label{fig:stream}
%\end{figure}

%\begin{table}[ht]
%\centering
%\begin{tabular}{|l|l|l|}
%\hline
%Condition & n & p \\
%\hline
%A & 5 & 0.1 \\
%\hline
%B & 10 & 0.01 \\
%\hline
%\end{tabular}
%\caption{\label{tab:example}Legend (350 words max). Example legend text.}
%\end{table}

%Figures and tables can be referenced in LaTeX using the ref command, e.g. Figure \ref{fig:stream} and Table \ref{tab:example}.

\end{document}